\documentclass[prl,aps, twocolumn, superscriptaddress]{revtex4-1}
\usepackage{amsmath,amssymb}
\usepackage{graphicx}
\usepackage{wasysym}
\usepackage{amsfonts}
\usepackage{bm}
\usepackage{enumerate}
\usepackage{color}
\usepackage[resetlabels]{multibib}
\usepackage{epstopdf}
\usepackage{latexsym}
\usepackage[breaklinks,colorlinks = true,linkcolor = red,urlcolor=cyan,citecolor=red]{hyperref}

\newcommand{\bea}{\begin{eqnarray}}
\newcommand{\eea}{\end{eqnarray}}
\newcommand{\be}{\begin{eqnarray}}
\newcommand{\ee}{\end{eqnarray}}
\newcommand{\bw}{\begin{widetext}}
\newcommand{\ew}{\end{widetext}}

\def\ket#1{{|#1\rangle}}

\def\pdag{\phantom\dag}

\begin{document}
\title{Filling-enforced  Kondo semimetals in two-dimensional non-symmorphic crystals}
\author{J. H. Pixley}
\email{jpixley@umd.edu}
\affiliation{Condensed Matter Theory Center and the Joint Quantum Institute,
Department of Physics, University of Maryland, College Park, Maryland 20742-4111 USA}
\author{SungBin Lee}
\email{sungbinl@uci.edu}
\affiliation{Department of Physics, Korea Advanced Institute of Science and Technology, Daejeon 305-701, Korea}
\affiliation{Department of Physics and Astronomy, University of California, Irvine, CA 92697 USA}
\author{B. Brandom}
\email{brandom@uci.edu}
\affiliation{Department of Physics and Astronomy, University of California, Irvine, CA 92697 USA}
\author{S. A. Parameswaran}
\email{sidp@uci.edu}
\affiliation{Department of Physics and Astronomy, University of California, Irvine, CA 92697 USA}

\date{\today}
\begin{abstract}
We study the competition between Kondo screening and frustrated magnetism on the non-symmorphic Shastry-Sutherland Kondo lattice at a filling of two conduction electrons per unit cell. A previous analysis of this model identified a set of gapless partially Kondo screened phases intermediate between the Kondo-destroyed paramagnet and the heavy Fermi liquid. Based on crystal symmetries, we argue that (i)~both the paramagnet and the heavy Fermi liquid are {\it semimetals} protected by a 
 glide symmetry; and (ii)~partial Kondo screening breaks the symmetry, removing this protection and allowing the partially-Kondo-screened phase to be deformed into a Kondo insulator via a Lifshitz transition.
{We confirm these results using large-$N$ mean field theory and then use non-perturbative arguments to derive}
 a generalized Luttinger sum rule constraining the phase structure of 2D non-symmorphic Kondo lattices beyond the mean-field limit.
\end{abstract}

\maketitle

\noindent{\it Introduction.---} 
The interplay between the Kondo effect and magnetism in heavy fermion materials is a paradigmatic setting for competing electronic order~\cite{ColemanBook}. In these {rare-earth} intermetallic compounds a lattice of local moments from {strongly correlated} $d$ or $f$ orbitals {can hybridize with itinerant conduction electrons to} form a heavy Fermi liquid (FL),  
with a `large' Fermi surface that incorporates {\it both} constituents. 
 Intermoment exchange {induced by the Ruderman-Kittel-Kasuya-Yosida
(RKKY) mechanism} can suppress 
Kondo screening in favor of magnetic order, and much theoretical and experimental effort has focused on studying the intervening quantum critical point~\cite{HeavyFermion-review,Coleman-2001, Si-2001,*Si-2003, Coleman-2005,Gegenwart-2008,SiSteglichReview,Si-2014}. Magnetic frustration adds complexity to this scenario~\cite{Si-2006,*Si2-2010,Coleman-2010} by introducing quantum fluctuations that  favor local-moment singlet formation over magnetic order~\cite{Coleman-1989}. 
Several unconventional phases result from this competition, including a metallic valence bond solid (VBS) in {Y}b{A}l$_3${C}$_3$~\cite{Khalyavin-2013}, partial magnetic order in CePd$_{1-x}$Ni$_x$Al~\cite{LohneysenCe,CePdAlnew}, and quantum criticality without tuning in CeRhSn~\cite{Tokiwa-2015}. Other proposed possibilities, such as fractionalized quantum spin liquids (QSLs)~\cite{SenthilVojtaSachdev}, remain experimentally elusive.

\begin{figure}
\includegraphics[width = \columnwidth]{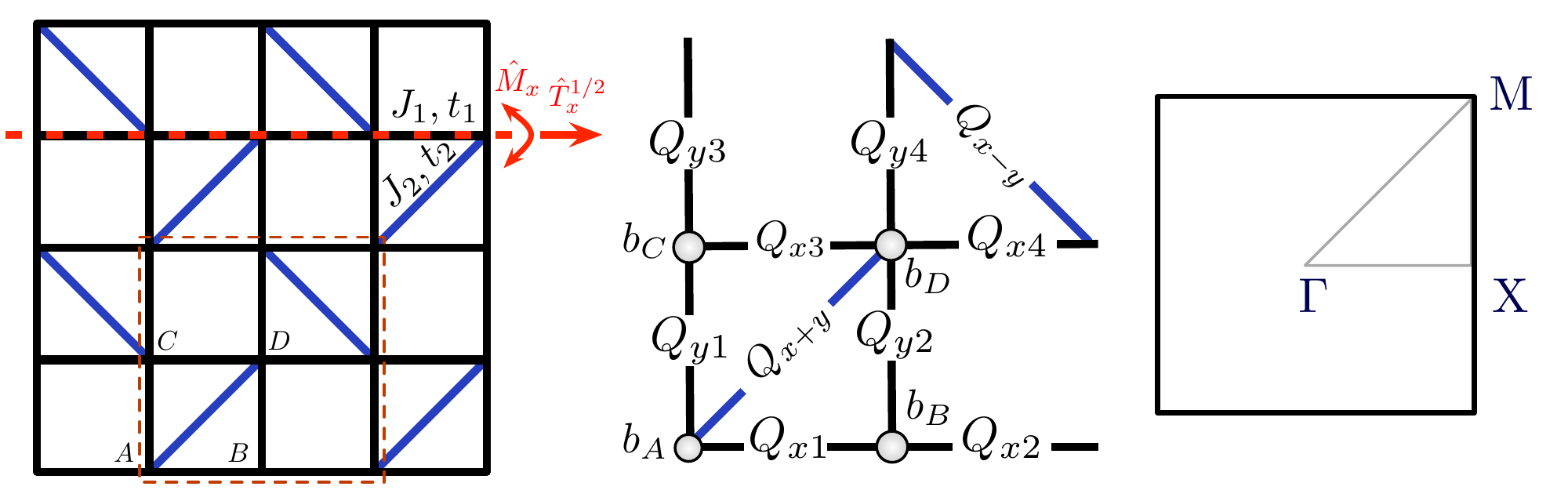}
\caption{\label{fig:lattice} SSL,  mean-field parameters, and reduced Brillouin zone. One of the glide planes is depicted, involving a reflection ($\hat{M}_x$) followed by a half-lattice-translation ($\hat{T}_x^{1/2}$).}
\end{figure}

A classic instance of {geometrical} frustration due to lattice structure
is furnished by the Shastry-Sutherland lattice~\cite{ShastrySutherland} (SSL; Fig.~\ref{fig:lattice}) relevant to a class of heavy fermion materials such as Yb$_2$Pt$_2$Pb, Ce$_2$Pt$_2$Pb, and Ce$_2$Ge$_2$Mg ~\cite{KimAronson1, *KimAronson2,*KimAronson3}. 
Recently, the phase structure of the Shastry-Sutherland Kondo lattice (SSKL) model was analyzed 
 using large-$N$ techniques and mean field arguments~\cite{Bernhard-2011,PixleySSL,Pixley-2015}, revealing several distinct phases. For weak Kondo coupling these include ordered antiferromagnets (at low frustration) and paramagnetic valence-bond solids (at strong frustration), whereas a heavy Fermi liquid phase was identified at strong Kondo coupling, {for a range of fillings}. 
Here, we re-examine these results with an emphasis on constraints imposed  on the phase structure by lattice symmetry, and focus on a filling of half an electron per site. 
 {At this filling} the four-site SSL unit cell contains exactly two conduction electrons ($\nu_c =2$) and 
---{since} there is a single local $f$-moment on each site--- four spins-$1/2$ ($N_s =4$). In the fully Kondo-screened phase, both of these must be included the Luttinger count --- the total number of electrons and local moments per unit cell, modulo those which can be incorporated into fully filled bands. This may be derived via Martin's periodic-lattice generalization of the Friedel sum rule~\cite{MartinSumRule}, or using topological arguments due to Oshikawa~\cite{OshikawaLuttinger}. Since $\nu \equiv \nu_c + N_s = 6$, the  Fermi surface of the Kondo-screened phase encloses zero net volume, as any even charge may be accommodated in filled, hybridized bands or in equal-volume electron and hole pockets. As we show below, the SSKL at $\nu_c=2$ is {\it gapless} for large Kondo coupling $J_K$. A naive expectation based on Luttinger's theorem is that the gaplessness is `accidental' and that the system can be driven insulating via a symmetry-preserving Lifshitz transition where the Fermi surface vanishes.

However, we demonstrate that this gapless Kondo-screened phase is a {filling-enforced~\cite{WatanabeFEGapless}  Kondo semimetal}, protected by a non-symmorphic glide symmetry (reflection combined with a half-lattice translation) of the SSL: it cannot become insulating without breaking this symmetry, or triggering fractionalization. This extends previous results for purely electronic systems~\cite{Parameswaran:2013ty, WatanabeFillingConstraints,WatanabeFEGapless,YoungKaneDSM2d,YoungetalKaneDSM3D} to Kondo lattices.  At intermediate $J_K$, glide symmetry is spontaneously broken, leading to a partially Kondo screened insulator (PKSI), where the hybridization between local moments and conduction electrons is modulated 
within the unit cell while preserving the translational symmetry of the SSL.
{The PKSI is thus distinct from conventional KIs that the SSL hosts at a filling of one electron per site, $\nu_c=4$~\cite{Pixley-2015}, that preserve all symmetries.}  Previous work~\cite{PixleySSL} found an intermediate-$J_K$ `partially Kondo screened'  gapless phase with similar broken symmetries at $\nu_c=2$; unlike the large $J_K$ Kondo semimetal, this may be adiabatically deformed to the PKSI.

We substantiate these claims within a large-$N$ mean-field study of the SSKL. We map the {finite temperature} phase diagram, and discuss transitions between the Kondo insulator and its proximate semimetals. We unify these results by identifying a generalized `Luttinger invariant' for Kondo lattice models at even integer filling, and close with a discussion of possible extensions.

{\it Model and $SU(N)$ Mean-Field Theory.---} The SSKL is described by the  Kondo-Heisenberg Hamiltonian, 
\bea
H= \!\!\sum_{(i,j),\sigma}t_{ij}(c_{i\sigma}^{\dag}c^{\pdag}_{j\sigma} + \mathrm{h.c.})  + \sum_{(i,j)}J_{ij}{\bf S}_i\!\cdot\!{\bf S}_j+J_K\sum_i {\bf s}_{i}\!\cdot\!{\bf S}_i.\nonumber\\
\eea
This describes  decoupled conduction electrons ($c_{i\sigma}$) that hop between sites with amplitude $t_{ij}$  and local spin-1/2 moments (${\bf S}_i$), coupled via an on-site antiferromagnetic Kondo coupling $J_K>0 $. Here, ${\bf s}_i = c_{i\alpha}^{\dag}(\bm{\sigma}_{\alpha\beta}/2)c^{\pdag}_{i\beta}$ is the conduction electron spin density and the sum on $(i,j)$ ranges over  nearest (NN; $t_1$, $J_1$) and next-nearest  (NNN; $t_2, J_2$) neighbors on the SSL. 

\begin{figure}
\includegraphics[width = 0.9\columnwidth]{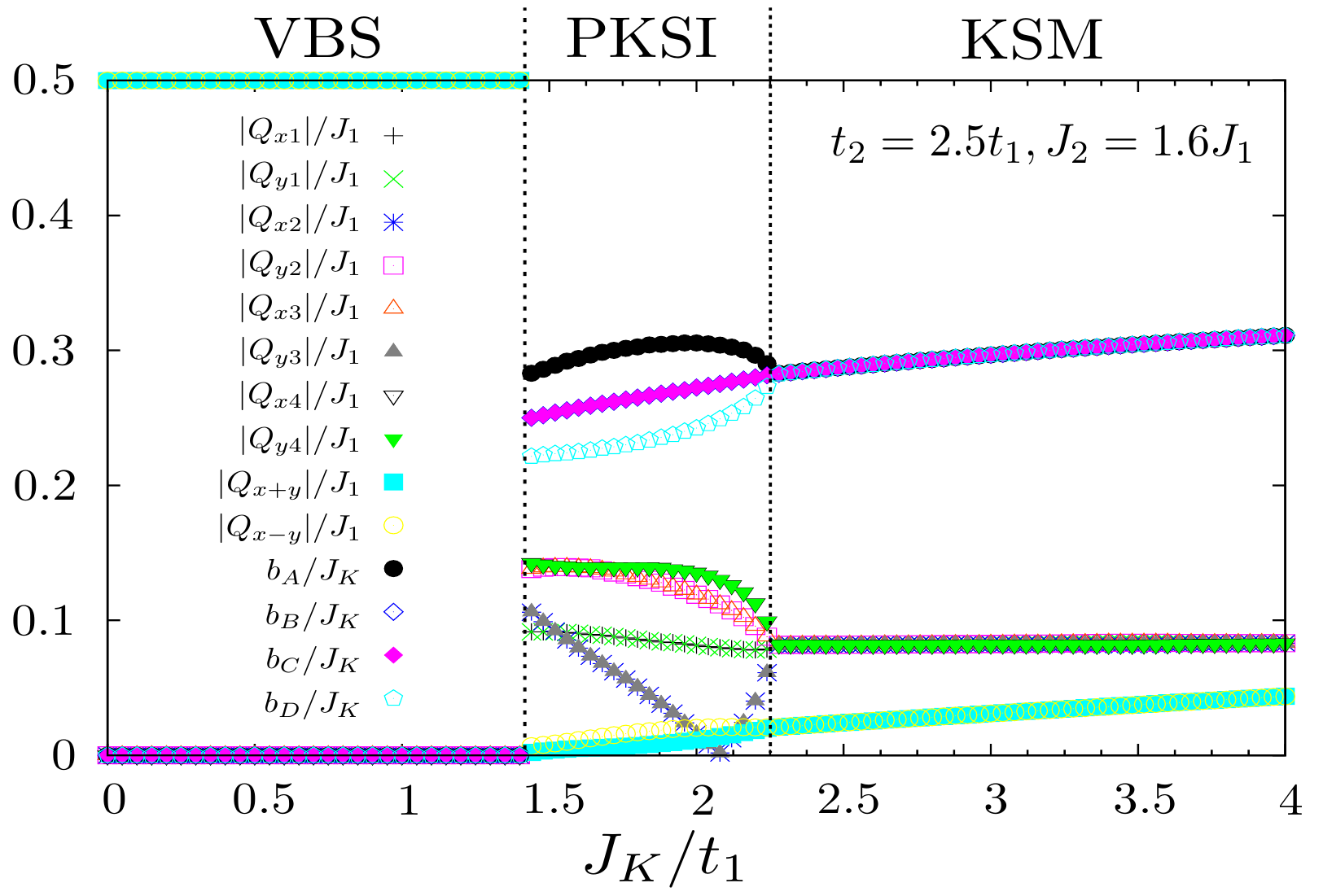}
\caption{\label{fig:MFTsol} Evolution of mean-field solutions
 with Kondo coupling $J_K$ at low temperature $T=t_1/100$. A VBS phase, where unscreened conduction electrons form a semimetal, exists for $J_K \lesssim 1.5 t_1$. The Kondo semimetal (KSM) found for $J_K \gtrsim 2.1 t_1$ cannot be trivially gapped while preserving glide symmetry. At intermediate $J_K$ glide is spontaneously broken, leading to a partially-Kondo-screened insulator (PKSI).}
\end{figure}

\begin{figure}
\includegraphics[width = \columnwidth]{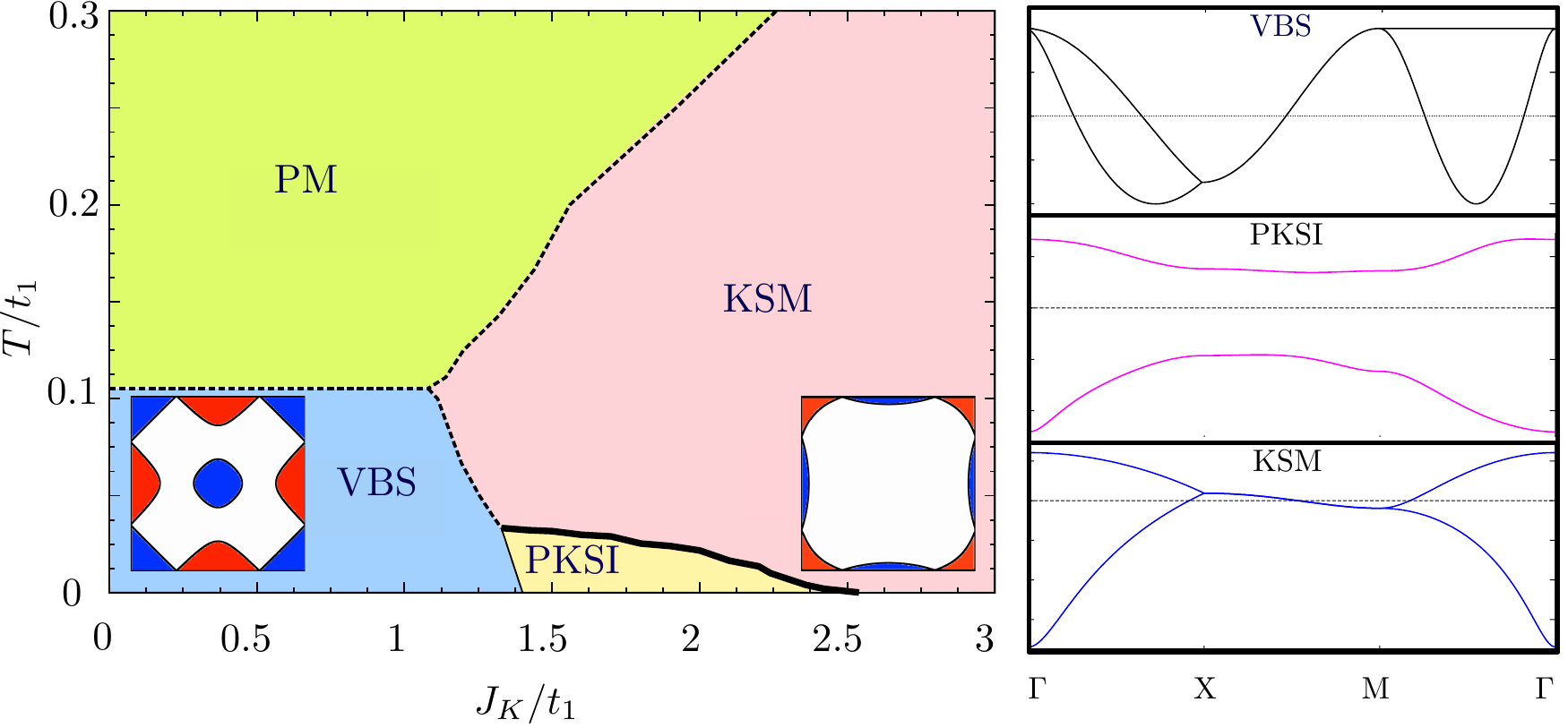}
\caption{\label{fig:pd} Large-$N$ Mean-field phase diagram of SSKL in the temperature-Kondo coupling ($T-J_K$) plane. Thick (thin) solid lines represent second- (first-)order transitions, while dashed lines are crossovers; the fermiology of the semimetallic KSM/VBS phases are inset, with electron (hole) pockets in red (blue). Right: band structures in the three $T=0$ phases. (only spin-degenerate bands nearest Fermi energy shown).  
}
\end{figure}

As a first step, we rewrite the model in terms of fermionic spinons ${\bf S}_i=f_{i\alpha}^{\dag}(\bm{\sigma}_{\alpha\beta}/2)f^{\pdag}_{i\beta}$, that are subject to the constraint $\sum_{\sigma}f_{i\sigma}^{\dag}f^{\pdag}_{i\sigma}=1$. We may then solve the problem via a large-$N$ mean-field approach, generalizing the spin symmetry from $SU(2)$ to $SU(N)$, so that saddle-point results become exact for $N\rightarrow \infty$.  Decoupling the Kondo and Heisenberg interactions via Hubbard-Stratonovich transformations that introduce the fields $b_i$ and $Q_{ij}$ respectively and allowing these fields to condense (i.e., acquire a non-zero saddle-point value), we arrive at the mean-field Hamiltonian~\cite{Coleman-1989,SenthilVojtaSachdev,PixleySSL}
\bea
H_{MF} &=& E-\sum_{(i,j),\sigma}( Q_{ij}^*f^{\dag}_{i\sigma}f^{\pdag}_{j\sigma} + \mathrm{h.c.} ) 
+\sum_{i,\sigma}\lambda_if^{\dag}_{i\sigma}f^{\pdag}_{i\sigma}
\label{eqn:Hmf}
\\
&+&\sum_{(i,j),\sigma}t_{ij} (c^{\dag}_{i\sigma}c^{\pdag}_{j\sigma} + \mathrm{h.c.}) 
-\sum_{i,\sigma}(b_i^*c^{\dag}_{i\sigma}
f^{\pdag}_{i\sigma} + \mathrm{h.c.}).
\nonumber
\eea
where
$E/N=\sum_i\left(|b_i|^2/J_K-\lambda_i/2\right)+
\sum_{(i,j)}|Q_{ij}|^2/J_{ij}$.  
Self-consistency requires that \bea\label{eq:selfcon}
b_i= \frac{J_K}{N}\sum_{\sigma}\langle c_{i\sigma}^{\dag}f^{\pdag}_{i\sigma} \rangle\,\,\text{and}\,\,
 Q_{ij}= \frac{J_{ij}}{N}\sum_{\sigma}\langle f^{\dag}_{i\sigma}f^{\pdag}_{j\sigma} \rangle,\eea
and we take $b_i$ to be real. We restrict ourselves to translationally-invariant mean-field solutions    
but do not enforce any symmetry within the unit cell. This permits us to access states that break lattice point-group symmetries but preserve translations. Such solutions may be parametrized in terms of 18 independent complex parameters: for each of the four sites in the unit cell  $\lambda_i$ enforces the constraint $\sum_{\sigma}\langle f_{i\sigma}^{\dag}f^{\pdag}_{i\sigma}\rangle=1$, while $b_i$ measures the hybridization on the site, 
and on each of the ten inequivalent bonds $Q_{ij}$ measures the strength of the singlet order (Fig.~\ref{fig:lattice}). Previous work~\cite{PixleySSL,Pixley-2015} has exclusively studied the regime $2t_1> t_2$;
 here we consider the opposite regime with $2t_1<t_2$ (we fix $t_2/t_1=2.5$ without loss of generality). As noted, we restrict ourselves to a filling of half an electron per site ($\nu_c =2$).  We numerically solve (\ref{eq:selfcon})
and track the soluton as a function of $J_K$; the results, shown in Figs.~\ref{fig:MFTsol} and~\ref{fig:pd}, are as follows. 

{
 {\bf (1)} The $J_K=0$ phase is stable for a finite range of $J_K\lesssim 1.5t_1$. In this case the only nonzero $Q_{ij}$ are $Q_{x +y} = Q_{x-y}$ so that spinon bands remain flat; meanwhile, the hybridization $b_i=0$ on all sites, so we can treat the spinon and conduction electron bands as decoupled.  
 Since $N_s = 4$, the lower pair of the four spinon bands are fully filled.  For $\nu_c =2$,  the chemical potential intersects the lower pair of the four total conduction electron bands. The spin-degenerate bands `stick'  in pairs due to glide symmetry~\cite{Parameswaran:2013ty, WatanabeFillingConstraints,WatanabeFEGapless,YoungKaneDSM2d,YoungetalKaneDSM3D} across the X-M face of the Brillouin zone (BZ; labeled as in Fig.~\ref{fig:lattice}) and cannot be detached  without breaking symmetry, though additional perturbations may reduce the sticking along X-M to nodes~\cite{supmat}. The resulting Fermi surface has an electron pocket centered at  X  and hole pockets centered at $\Gamma$, M,  that  enclose zero net charge: it is a semimetal. As the singlet bonds are identical to those of the pure Heisenberg model~\cite{ShastrySutherland}, we label this the VBS phase; this preserves all symmetries of the SSL, including the glide symmetry responsible for the band sticking.}
 
{
 {\bf (2)} Turning next to large $J_K\gtrsim2.1 t_1$, we find a Kondo-screened phase with $\lambda_i =\lambda$ and $b_i =  b \neq0$ on all sites of the unit cell, and $Q_{xi} = Q_{yi}$ also nonzero and distinct from $Q_{x+y} = Q_{x-y}$. This solution also preserves all symmetries of the SSL. The spinon and conduction electron bands are hybridized and cannot be considered separately; as glide symmetry remains unbroken, the spin-degenerate hybridized bands are again stuck in pairs along the X-M face, as in the VBS phase. The bands near the Fermi energy 
 split linearly at X and quadratically at M  (Fig.~\ref{fig:pd}), leading to hole and electron pockets centered at these points in the BZ.  For filling  $\nu_c + N_s = 6$  the chemical potential intersects the second-lowest pair of hybridized bands, again leading to a (compensated) semimetal. We dub this phase a Kondo semimetal (KSM). Even though it is gapless and emerges when Kondo screening dominates magnetism,  the KSM is distinct from a heavy Fermi liquid, since its Luttinger count vanishes:  as in {the} VBS {phase}, the  electron and hole Fermi surfaces enclose zero net charge (Fig.~\ref{fig:pd}).}

 {\bf (3)} At intermediate Kondo coupling, the mean-field solution no longer preserves glide symmetry: for instance, the hybridization is modulated within the unit cell $b_A \neq b_B = b_C \neq b_D$, and so is not invariant under the pairwise exchange $A\leftrightarrow B$, $C\leftrightarrow D$ generated by acting with the glide on a translationally invariant state. Similarly, the singlet pattern breaks the glide symmetry. As a consequence of the broken symmetry the hybridized bands no longer stick pairwise, and therefore at the filling $\nu_c +N_s=6$ we see that the Fermi surface lies in a gap (Fig.~\ref{fig:pd}). As the system screens unequally on different sublattices and is gapped, we identify this as a partially Kondo screened insulator (PKSI). This is, to our knowledge, the first example of a Kondo insulator where  screening spontaneously breaks lattice symmetry; absent symmetry breaking the only other route to opening a gap is to trigger topological order, a case we do not consider here.  It may be possible to probe glide symmetry breaking via scattering experiments, where it is signaled by the reappearance of spectral weight at Bragg peaks `systematically extinguished'  by the glide symmetry.

Previous studies of this regime~\cite{PixleySSL} with $2t_1>t_2$ found partially-Kondo-screened gapless phases  {for $0<\nu_c<4$}. For $\nu_c=2$, such a phase is now identified as an `accidental' semimetal with broken glide symmetry and electron and hole pockets enclosing zero net volume. This phase is obtained by  driving the PKSI through  a Lifshitz transition so that its bands intersect the Fermi energy. We  show below how  glide symmetry can distinguish such accidental semimetals from the filling-enforced KSM and VBS semimetals via a generalized Luttinger sum rule.

{\it Finite-Temperature Phase Diagram.---} {
 Usually, Kondo screening onsets via a crossover, as no broken symmetry distinguishes the $T\rightarrow 0$ heavy Fermi liquid from the high-temperature paramagnet (PM). A similar argument applies to the VBS state, since the valence bond pattern preserves symmetry (magnetic order would require a true phase transition). Accordingly, we identify  the PM-VBS and PM-KSM lines as crossovers,  although they appear as transitions within mean-field theory (Fig.~\ref{fig:pd}).
The intermediate PKSI phase, however, breaks a discrete glide symmetry, and hence can form via a finite-temperature  transition in $d=2$. Note that the transition temperature $T_c$ is an order of magnitude lower than the screening scale $T_K$; hence, the relevant degrees of freedom are hybridized fermions rather than bare electrons (typically, $T_K\sim 10$K so $T_c$ is experimentally accessible).
At mean-field level, the PKSI-VBS transition appears first order, though this may be an artifact of the large-$N$ approach. The $T=0$ PKSI-KSM transition appears continuous within mean-field theory, and is worthy of further study~\cite{BPLP-unpub}.}

{\it Generalized Luttinger Invariant.---} The SSKL illustrates general symmetry constraints for non-symmorphic Kondo lattices in $d=2$, that we now derive. Let us consider a lattice of $L_x\times L_y$ 
unit cells  with periodic boundary conditions  (i.e. a torus) described by a non-symmorphic space group $\mathcal{G}$. $\mathcal{G}$ includes at least one
glide reflection  ${G}_x$, 
 involving a mirror reflection $M_x$ about the $x$ axis followed by a translation through a half-lattice vector in the mirror plane:
$G_x =  T_x^{1/2}M_x: (x,y) \rightarrow (x+1/2, -y)$.  
We restrict ourselves to 
 Hamiltonians with $SO(3)$ spin rotation symmetry so that we can separately couple to ``up'' and ``down'' spins relative to a fixed magnetization axis and take reflections to act trivially on spin. 
A single quantum $\Phi_0$ of Aharonov-Bohm flux that couples only to one spin species is threaded through one of the non-contractible loops of the torus, e.g by adiabatically increasing a uniform time-dependent vector potential from $A_x=0$ to $A_x = \frac{2\pi}{L_x}$ (we take $\hbar=e=1$ so $\Phi_0=2\pi$). 
Under this process, the ground state $\ket{\Psi_0}$ evolves into a (possibly distinct) state $\ket{\Psi_0'}$; however,  the Hamiltonian with the vector potential $A_x$ commutes with the glide symmetry operator at all times so that both states share the same glide eigenvalue, $G_x \ket{\Psi_0} =  e^{i g_0} \ket{\Psi_0}, G_x \ket{\Psi_0'} =  e^{i g_0 } \ket{\Psi_0'}$. However, $\ket{\Psi'_0}$ is an eigenstate of a Hamiltonian $H(\Phi_0)$ with an inserted flux; we may return to $H(0)$ by implementing a large gauge transformation, so that $\ket{{\Psi}'_0}\rightarrow \ket{\tilde{\Psi}_0} = U_{\sigma}\ket{\Psi'_0}$  ($\sigma= \uparrow, \downarrow$ denotes spin), with
\be
U_{\sigma} = \exp\left[\frac{2\pi i}{L_x} \sum_{\vec r} x (n^{\sigma}_{c,\vec{r}} + S^z_{\vec r}) \right],
\ee
where the second term is required to keep the Kondo coupling invariant. It is straightforward to show that $U_{\sigma}^{-1} T_x U_{\sigma}  = T_x \exp\left[ 2\pi i \{n^\sigma_c + N_s(\frac{1}{2} \pm m)\}L_y \right]$ where 
the  contribution proportional to $N_s/2$ is from a boundary term, $m$ is the magnetization per localized spin, and we  take $+ (-)$ for $\sigma = \uparrow (\downarrow)$~\cite{OshikawaLuttinger}. Since $G_x =  T_x^{1/2}M_x $ and $M_x$ leaves the magnetization axis unchanged, it follows  that $U_{\sigma}^{-1} G_x U_{\sigma} = G_x \exp\left[i \pi  \{n^\sigma_c + N_s(\frac{1}{2} \pm m)\}L_y \right]$. Thus,  $G_x\ket{\tilde\Psi_0} = e^{ig} \ket{\tilde\Psi_0}$ with
\be\label{eq:UVcounting}
e^{i(g- g_0)} =  e^{i\pi[n^\sigma_c + N_s(\frac{1}{2} \pm m)]L_y}.
\ee
Now, let us assume that the system is described by Fermi liquid theory. Flux insertion corresponds to shifting every quasiparticle excitation via $k_x \rightarrow k_x +2\pi/L_x$, producing quasiparticles and quasiholes on opposite sides of the Fermi surface. The net result of this is equivalent (e.g., by applying Stoke's theorem in the BZ) to a shift of all the filled states by $2\pi/L_x$, and therefore the net change in momentum of the system in the adiabatic process is given by $\Delta P_x = \frac{2\pi}{L_x} N^{(L)}_{F,\sigma}$, where $N^{(L)}_{F,\sigma}$ is the total number of filled spin-$\sigma$ states {in the finite size system}. Since the glide involves a half-translation but does not mix different spin projections, the  change in the glide quantum number is exactly one-half the momentum change,
\be\label{eq:IRcounting}
e^{i(g -g_0)} = e^{i\frac{\pi}{L_x} N^{(L)}_{F,\sigma}}.
\ee
Comparing Eqs. \eqref{eq:UVcounting} and \eqref{eq:IRcounting} and setting $L_x =L_y =L$, we find that $(\chi^{(L)}_{F,\sigma} - \nu_\sigma)L = 2p$ where $p$ is an integer, and we defined $\nu_\sigma \equiv \nu^\sigma_c + N_s(\frac{1}{2} \pm m)$ and $\chi^{(L)}_{F,\sigma} \equiv  N^{(L)}_{F,\sigma}/L^2$. A consistent thermodynamic limit for $L$ odd then requires
\be
\chi_{F}^{\sigma}  \equiv  \nu_\sigma (\text{mod 2})
\ee
where $\chi_{F}^{\sigma}  \equiv \lim_{L\rightarrow\infty} \chi^{(L)}_{F,\sigma}$ is the new (spin filtered) Luttinger invariant.
A similar computation with $T_x$ replacing $G_x$, can be used to constrain the Fermi volume, via $\frac{V^\sigma_F}{(2\pi)^2} =\nu_\sigma$ (mod 1) ~\cite{OshikawaLuttinger}. Let us now examine the behavior of these invariants in the spin symmetric case, where $m=0$, $\nu_c^\uparrow= \nu_c^\downarrow = \frac{\nu_c}{2}$, $V_F =  V_F^\uparrow = V_F^\downarrow$, and $\chi_F^\uparrow =\chi_F^\downarrow$. Consider a filling $\nu =  4p+2$, where $p$ is an integer   (as in the example above); then, $\nu_\sigma = \frac{\nu}{2} = 2p+1$, and
\be\label{eq:invariants}
V_F^\sigma =0\,\,\,\, \text{and} \,\,\,\, \chi_F^\sigma = 1,
\ee
i.e., the Fermi volume vanishes, while the  generalized Luttinger invariant is non-zero.  

The non-zero Luttinger invariant indicates a nontrivial spectral flow (reflected by the change in glide quantum numbers) under flux insertion~\cite{NSLuttinger}, which can be satisfied either by the presence of  gapless excitations of the ground state or by the existence of a fractionalized topological quasiparticle. A Kondo insulator  --- which is a gapped, non-fractionalized phase --- cannot respond to the insertion of a flux by changing its glide quantum number; hence, it cannot {have a nonzero Luttinger invariant}.  
{However, the Fermi volume is zero [Eq.(\ref{eq:invariants})]}. As we do not consider fractionalized phases, the only possibility consistent with these two requirements  is for the system to be a semimetal with band crossings protected by glide symmetry. As long as glide symmetry is preserved, the electron and hole Fermi pockets can be shrunk to point nodes but cannot be completely removed --- as in the symmetric phases (VBS, KSM) identified in our study. Breaking glide symmetry allows $\chi_F^\sigma =0$ permitting a gapped phase (as in PKSI). A modification of this argument was presented for the $N_s=0$ case in~\cite{NSLuttinger}. This generalized Luttinger sum rule may be `topologically enriched' by allowing for the possibility of gapped, symmetry-preserving phases with fractionalized quasiparticles~\cite{SenthilVojtaSachdev,LuttingerExtension,EnrichedLuttinger,BPLP-unpub}.

{\it Concluding Remarks.---} We have examined the role of glide symmetries in determining the phase structure of a canonical 2D non-symmorphic Kondo lattice, the SSKL, and  identified a filling-enforced Kondo semimetal. We have also demonstrated that competition with frustrated magnetism can lead to a broken-symmetry Kondo insulator. While we use a large-$N$ approximation, our results are  consistent with a non-perturbative Luttinger sum rule that applies well away from the mean-field limit. Our symmetry analysis  provides a unified perspective on the Doniach diagram~\cite{DONIACH1977231} of 2D nonsymmorphic Kondo lattices. For fillings $\nu = 4p+2$, corresponding to vanishing Fermi volume, the nonzero Luttinger invariant requires that any symmetry-preserving phase either remains gapless or else has topological order. The former possibility --- a symmetric semimetal --- is likely at large and small $J_K$, where either magnetism or Kondo screening dominates. At intermediate coupling, competition leads to the opening of a gap; absent topological order, such a gapped phase must necessarily break glide symmetry, as in the PKSI we find here. For $\nu=4p+4$~\cite{Pixley-2015}, both the Fermi volume and the generalized invariant vanish and these constraints do not apply. {Although spin-orbit coupling (SOC) is challenging to treat using flux insertion, if we apply existing results~\cite{WatanabeFillingConstraints,WatanabeFEGapless,YoungKaneDSM2d} on filling-enforced semimetals to the hybridized bands at these fillings our results remain unchanged if time-reversal symmetry is present. There is then also the additional interesting possibility that the PKSI may be a topological Kondo insulator, as topological insulators can emerge naturally from  filling-enforced SOC semimetals upon breaking glide symmetry~\cite{WatanabeFEGapless,YoungKaneDSM2d,FKMDiamond}.}  As glide is the only non-symmorphic symmetry in $d=2$, this exhausts possible non-fractionalized symmetric phases at large- and small-$J_K$ for 2D Kondo lattices at commensurate filling {(i.e., 
$V_F=0$)}. {Our work suggests that non-symmorphic lattices are natural hosts for strongly correlated semimetals and descendant phases; in the future, we hope to extend our analysis to all 157 non-symmorphic 3D space groups~\cite{BPLP-unpub}.}

{\it Acknowledgements.---}
We are grateful to A. Vishwanath and P. Bonderson for helpful correspondence, M. Zaletel for discussions, and A. Vishwanath, R. Yu and M. Zaletel for comments on the manuscript. J.H.P. thanks Rong Yu, Silke Paschen, and Qimiao Si for collaborations on related work. This work was supported by   JQI-NSF-PFC, LPS-MPO-CMTC
and Microsoft Q (J.H.P.), KAIST start-up funds (S.B.L.), UC Irvine start-up funds (S.A.P., S.B.L.) and NSF Grant No. DMR-1455366 (S.A.P.). J.H.P. and S.B.L. acknowledge the hospitality of the Aspen Center for Physics (NSF Grant no. PHY-1066293)  and J.H.P. acknowledges the hospitality of UC Irvine.

\bibliography{kondo_bib}

\clearpage

\onecolumngrid
\setcounter{figure}{0}
\makeatletter
\renewcommand{\thefigure}{S\@arabic\c@figure}
\setcounter{equation}{0} \makeatletter
\renewcommand \theequation{S\@arabic\c@equation}
\renewcommand \thetable{S\@arabic\c@table}

\section*{{\Large Supplementary Material}}
\section{Solving the large $N$ mean field equations}
We minimize the free energy at a temperature $T$ with respect to the self consistent mean field parameters $Q_{ij}$, $b_X$, and $\lambda_X$ by solving the non-linear mean field equations [Eq. (3) the main text]. To solve the equations for a free energy minimum we use Broyden's mixing method~\cite{Broyden65,NumRecp07,Johnson88}.
The method solves the self consistent set of equations $\boldsymbol{X}=\boldsymbol{Y}[\boldsymbol{X}]$ iteratively starting from an initial guess of the ${\bf X}={\bf X}^{(0)}_{\mathrm{in}}$. At the $n$-th iteration a new set of values of the ${\bf X}$'s are obtained via
\begin{equation}
 \boldsymbol{X}^{(n)}_{\mathrm{out}} = \boldsymbol{Y}[\boldsymbol{X}^{(n)}_{\mathrm{in}}].
\end{equation}
The input  $\boldsymbol{X}^{(n+1)}_{\mathrm{in}}$ for the next iteration is given by
\begin{equation}
 \boldsymbol{X}^{(n+1)}_{\mathrm{in}} = \boldsymbol{X}^{(n)}_{\mathrm{in}} - \boldsymbol{B}^{(n)} (\boldsymbol{X}^{(n)}_{\mathrm{out}} - \boldsymbol{X}^{(n)}_{\mathrm{in}}),
\end{equation} 
here $\boldsymbol{B}^{(n)}$ is an approximation to the inverse of the Jacobian matrix of the non-linear equations at the $n$-th iteration. Thus, Broyden's method provides a way to approximate the inverse Jacobian $\boldsymbol{B}^{(m)}$ at each iteration such that the self consistent solution is obtained in a numerically fast manner~\cite{Broyden65,NumRecp07,Johnson88}. To find the global minimum of the free energy we perform a random search over the initial ${\bf X}^{(0)}_{\mathrm{in}}$'s, (we used $10^3$ different initial guesses) and take the solution with the lowest free energy. Once the correct phases are identified, we determine the phase boundaries by using the final solutions as initial guesses in the mixing procedure to bias the solution across each respective transition.

\section{Discussion of Band-Touchings for the Non-Symmorphic Lattice}
We briefly discuss the band contacts along high-symmetry lines in our model, that result from glide symmetry. For more details, we refer the reader to ~\cite{YoungKane}; we closely follow their line of argument in what follows. We are primarily interested in the space group of the Shastry-Sutherland lattice: the wallpaper group (as 2D space groups are termed) $p4g$. The key symmetries of interest are the glide planes; one of these is depicted in Figure 1 of the main text, and the other is related by a $\pi/2$-degree rotation. Note that while the $p4g$ space group appears to have additional glide planes oriented at $\pi/4$ with respect to the ones identified, those can be rewritten as the product of a reflection and an ordinary (rather than fractional) lattice translation, and as such do not lead to any protected degeneracies~\cite{SPNSthm}. A glide plane involves a mirror reflection followed by a lattice translation: for example, for the glide shown in the text  $G_x \equiv \{g|\mathbf{t}\}$ where $g= M_x: (x,y) \rightarrow (x,-y)$ is a reflection about the $x$-axis, and $\mathbf{t}$ is a half-translation along the mirror axis $\mathbf{t}: (x,y) \rightarrow (x+1/2, y)$ (we take the spacing between unit cells to be unity). Note that this operation leaves the $x$ axis invariant, so that we have in reciprocal  space that $g\mathbf{k} = \mathbf{k}$ for $\mathbf{k} =(k_x,0)$. In other words,  we may choose Bloch eigenstates along the invariant space (the $\Gamma$-X line $(k_x,0)$ in momentum space) to also be eigenstates of the  glide, so that along this line we have
\begin{equation}
G_x |{u_{{k_x,0}}^{\pm}}\rangle = \pm \lambda e^{i{k_x/2}} |{u_{{k_x,0}}^{\pm}}\rangle 
\end{equation}
where $\lambda^2 = M_x^2$ and can be determined at $k_x =0$. Now suppose we send $(k_x,0) \rightarrow (k_x +2\pi, 0)$; we see that the two eigenstates must switch places~\cite{SPNSthm}, so that absent other degeneracies the bands must cross an odd number of times as they cross the BZ~\cite{YoungKane}. 

So far we have only used the glide symmetry. In the models we study, time reversal  (denoted $\Theta$) is also a symmetry. As we have no spin-orbit coupling, we may take $\Theta^2 = +1$, $M_x^2=+1$, so that $\lambda = \pm 1$. Then, we see that the crossing occurs at the BZ boundary, i.e. at $\text{X} = (\pi,0)$, since at this point the glide eigenvalues $\pm i$ are exchanged by complex conjugation (and hence by $\Theta$.) From this, we see that the bands stick in pairs, so that at even filling $4n+2$ (for spinful electrons) the chemical potential will always intersect such a stuck-together pair, leading to semimetallic behavior.

Note that the models we study may have additional `accidental' symmetries owing to the minimal set of tight-binding parameters $t_1, t_2$, that we have used. Such additional symmetries are responsible for the sticking along the entire X-M face rather than only at the high-symmetry points (as in the discussion in~\cite{YoungKane}). Additional, symmetry-allowed perturbations, or spin-orbit interactions, can gap these out, so that rather than have line stickings along the BZ face, we have only point nodes.

\end{document}